\documentclass[
    ,final            
  ,amsmath,amssymb]
  {aipproc}
\usepackage{amssymb}
\layoutstyle{6x9}

\begin{document}

\title{Light Fermionic Dark Matter and its Possible Detection in Neutrino Experiments}



\author{Jennifer Kile}{
  address={Brookhaven National Laboratory, Upton, NY 11973}
}


\begin{abstract}
We explore the potential for the direct detection of light fermionic dark matter in neutrino detectors.  We consider the possible observation of the process $\bar{f} p \rightarrow e^+ n$, where $f$ is a dark matter fermion, in a model-independent manner.  All operators of dimension six or lower which can contribute to this process are listed, and we place constraints on these operators from decays of $f$ which contain $\gamma$ rays or electrons.  One operator is found which is sufficiently weakly constrained that it could give observable interactions in neutrino detectors.  We find that Super-Kamiokande can probe the new physics scale for this operator up to $O(100\mbox{ TeV})$.
\end{abstract}

\maketitle


\section{Introduction}
Despite years of effort, little is known of the nature of dark matter (DM).  We do not know how many species of DM exist, what their masses are, how they interact with each other, or how they interact with Standard Model (SM) particles.  Given this lack of knowledge of the properties of DM, we choose a possible interaction between DM and SM particles with a distinctive experimental signature and study it using a model-independent approach.  We perform an effective-operator analysis of the interaction $\bar{f} p \rightarrow e^+ n$ and its observability in neutrino detectors.  This talk is based largely on \cite{Kile:2009nn}.  

In the usual DM direct detection scenario, a DM particle is taken to scatter elastically off of some SM particle, such as an atomic nucleus.  In this case, the momenta of the final-state particles are of the order of the momentum of the incoming DM particle.  Instead, the interaction we consider is inelastic; we take $m_f>>m_e$.  In this case, the energy of the outgoing $e^+$ is very close to $m_f$.  Thus, the experimental signature of this interaction is similar to that of the well-studied SM neutrino interaction $\bar{\nu}_e p \rightarrow n e^+$, except that the $e^+$ is essentially monoenergetic.  If $m_f\sim 1-100$ MeV, then we could expect that existing solar and reactor neutrino detectors may be able to observe this DM interaction.  Although we do not assume that $m_f$ is in this range, we find, after insisting that $f$ be long-lived and decay sufficently slowly into easily observable final states containing $\gamma$ rays and $e^+e^-$, that the least constrained range for $m_f$ is $\lesssim O(100\mbox{ MeV})$.

We now discuss the basis of operators which can contribute to  $\bar{f} p \rightarrow e^+ n$.

\section{Operator Basis}
Here, we list our operator basis.  We take $f$ to be a fermion and an SM singlet.  We consider operators up to dimension six which are invariant under the SM gauge group $SU(3)\times SU(2)\times U(1)$; we eliminate redundant operators using integration by parts and the equations of motion for the fields.  We exclude operators which contribute at tree level to neutrino mass, as these are tightly constrained.  With these criteria, we obtain 
\begin{eqnarray}
{\cal O}_{W} &=& g \bar{L} \tau^a \tilde{\phi} \sigma^{\mu\nu} f W^a_{\mu\nu}\nonumber\\
{\cal O}_{\tilde{V}} &=& \bar{\ell}_R\gamma_{\mu}f \phi^{\dagger} D_{\mu} \tilde{\phi}\nonumber\\
{\cal O}_{VR} &=& \bar{\ell}_R \gamma_{\mu} f \bar{u}_R \gamma^{\mu} d_R, \\
{\cal O}_{Sd} &=& \epsilon_{ij }\bar{L}^i  f \bar{Q}^j d_R\nonumber\\
{\cal O}_{Su} &=& \bar{L} f \bar{u}_R Q \nonumber\\
{\cal O}_{T} &=& \epsilon_{ij }\bar{L}^i \sigma^{\mu\nu} f \bar{Q}^j \sigma_{\mu\nu} d_R \nonumber
\end{eqnarray}
where $L$ and $Q$ are the SM $SU(2)$ lepton and quark doublets,  $\ell_R$, $u_R$, and $d_R$ are the SM $SU(2)$ singlets, $\phi$ is the SM Higgs field, and $\tilde{\phi}= i \tau^2 \phi^*$.  All of the operators ${\cal O}_I$ are dimension-six and thus suppressed by $\Lambda^2$, where $\Lambda$ is a new physics scale, and each of the ${\cal O}_I$ is multiplied by a coefficient $C_I$.  In all of these operators, $f$ is right-handed.

We now constrain these operators using DM lifetime and decays.
 
\section{Constraints from Dark Matter Lifetime}
We now place limits on the above operators by insisting that $f$ have a lifetime at least as long as the age of the universe and that it decay to easily observable SM final states sufficiently slowly to have not yet been detected.

The operator ${\cal O}_{W}$ gives the tree-level decay $f\rightarrow \nu \gamma$.  DM decays to a two-particle final state which contains a photon are constrained \cite{Yuksel:2007dr} to occur on a timescale $\gtrsim O(10^{26}\mbox{ s})$ for DM masses between $\sim 1$ MeV and $\sim 100$ GeV.  From these limits, we obtain 
\begin{equation}
\frac{|C_W|^2}{\Lambda^4} \lesssim \frac{1}{(8 \times 10^7 \mbox{ TeV})^4} \left(\frac{1\mbox{ MeV}}{m_f}\right)^3
\end{equation}
which is far beyond the reach of present-day terrestrial experiments.

${\cal O}_{\tilde{V}}$ gives a tree-level decay $f\rightarrow e^+ e^- \nu$.  DM decays containing $e^+e^-$ are constrained \cite{Picciotto:2004rp} to occur on a timescale $\gtrsim 10^{17}$ years for values of $m_f$ relevant here.  Thus, we obtain
\begin{equation}
\frac{|C_{\tilde{V}}|^2}{\Lambda^4}\lesssim \frac{1}{(1-4\times 10^6\mbox{ TeV})^2}
\end{equation}
where the range of values corresponds to the range of $m_f$ which will be relevant for Super-Kamiokande, $20\mbox{ MeV }<m_f<80\mbox{ MeV}$.  Again, the new physics scale for this operator is constrained to be far beyond what is accessible at terrestrial experiments, and is more strongly constrained for larger values of $m_f$.  We will use the limits on these two operators to get limits on the others via operator mixing.

If $m_f>m_{\pi} + m_e$, ${\cal O}_{VR}$ will give the tree-level decay $f\rightarrow \pi^+ e^-$.  Therefore, for this operator, we only consider the low-mass region, $m_f\lesssim m_{\pi}$.  At one-loop order, ${\cal O}_{VR}$ mixes into ${\cal O}_{\tilde{V}}$ and thus gives the decay $f\rightarrow e^-e^+\nu_e$.  However, the diagram for this mixing is suppressed by both the $u$ and the $d$ Yukawa couplings.  This suppression greatly weakens the lower bound on the new physics scale for this operator.  We find
\begin{eqnarray}
\frac{|C_{VR}|^2}{\Lambda^4} & \lesssim & \frac{1}{(20\mbox{ TeV})^4} \mbox{ ($m_f=20$ MeV)}\nonumber\\
& \lesssim &\frac{1}{(50 \mbox{ TeV})^4} \mbox{ ($m_f=50$ MeV)}\label{eq:ovr15}
\\
& \lesssim &\frac{1}{(80\mbox{ TeV})^4} \mbox{ ($m_f=80$ MeV)}.\nonumber
\end{eqnarray}

As the one-loop mixing into ${\cal O}_{\tilde{V}}$ does not accurately represent contributions from loop momenta below a few hundred MeV, we also consider the diagram where $f$ decays to $e^-e^+\nu_e$ via a virtual $\pi^+$.  This diagram, however, gives weaker constraints on the new physics scale, ranging between a few and $50$ TeV.

Given the relatively weak limits from DM decays on ${\cal O}_{VR}$, we consider constraints from ${\cal O}_{VR}$ mixing into ${\cal O}_W$, $f-\nu$ mixing (which gives $f\rightarrow e^-e^+\nu_e$ and $f\rightarrow \nu\nu\bar{\nu}$), and $\pi^+$ decay.  These limits are not competitive with that from mixing into ${\cal O}_{\tilde{V}}$.

Finally, we constrain ${\cal O}_{Sd}$, ${\cal O}_{Su}$, and ${\cal O}_{T}$ through their mixing into ${\cal O}_W$.  For these operators, this mixing is suppressed by only one small Yukawa coupling, and, thus the lower bound on the new physics scale for these operators is $>10^3$ TeV.

For the rest of this work, we consider only the most weakly constrained operator, ${\cal O}_{VR}$.

\section{Signatures in Neutrino Experiments}
In order to evaluate the observability of the interaction $\bar{f} p \rightarrow e^+ n$ in neutrino detectors, we must have some idea of the possible $\bar{f}$ flux at the Earth's surface.  As the mass density of DM is thought to be of order  $\sim 0.3 \mbox{ GeV}/\mbox{cm}^3$ \cite{Caldwell:1981rj}, and its velocity $v_f$ relative to the Earth is believed to be approximately $O(10^{-3})$ (for $c=1$) \cite{Kamionkowski:1997xg}, we can estimate the $\bar{f}$ flux $\Phi_{\bar{f}}$ for the case where $\bar{f}$ comprises all of DM.  We obtain
\begin{equation}
\Phi_{\bar{f}}\sim\frac{0.3\mbox{ GeV}/\mbox{cm}^3}{m_f} v_f c \sim (10^{10},10^9,10^8) /\mbox{cm}^2\mbox{s}
\end{equation}
for $m_f=(1, 10, 100)$ MeV, respectively.  We can then compare this with the limit on the relic supernova $\bar{\nu}_e$ flux for $19.3 \mbox{ MeV} < E_{\nu} \lesssim 80 \mbox{ MeV}$ from Super-K \cite{Malek:2002ns} of $1.2$  $\bar{\nu}_e / \mbox {cm}^2 \mbox{s}$. 

We then calculate the ratio of the $\bar{f} p \rightarrow e^+ n$ cross-section to that of the SM neutrino interaction $\bar{\nu}_e p \rightarrow e^+ n$.  We ignore the proton-neutron mass difference, the electron mass, and any possible second-class currents.  This ratio must be smaller than the ratio of the above fluxes; we obtain 
\begin{equation}
\frac{|C_{VR}|^2 v^4}{8 |v_f| \Lambda^4}\leq \frac{1.2/\mbox{cm}^2\mbox{s}}{(0.3\mbox{ GeV}/\mbox{cm}^3) |v_f| c/m_f},
\end{equation}
where $v$ is the Higgs vacuum expectation value.  This gives
\begin{eqnarray}
\label{eq:results}
\frac{|C_{VR}|^2}{\Lambda^4} & \lesssim & \frac{1}{(120\mbox{ TeV})^4} \mbox{ ($m_f=20$ MeV)}\nonumber\\
& \lesssim &\frac{1}{(90 \mbox{ TeV})^4} \mbox{ ($m_f=50$ MeV)}\\
& \lesssim &\frac{1}{(80\mbox{ TeV})^4} \mbox{ ($m_f=80$ MeV)}.\nonumber
\end{eqnarray}
The scale probed will of course be lower if $\bar{f}$ comprises only a fraction of DM.

\section{Conclusions}
As our understanding of DM is far from complete, ``nontraditional'' interactions between DM and SM particles should not be ignored.  Here, we have considered the relevance of neutrino detectors to one such possible interaction,  $\bar{f} p \rightarrow e^+ n$, in a model-independent manner.  The main conclusions from this work are twofold.  First, the inelasticity of this interaction was critical to the possible direct detection of MeV-scale DM; this may indicate additional possibilities for the detection of light DM in existing neutrino or DM detectors.  Second, we found that the reach of neutrino detectors for observing light DM is potentially very impressive; new physics scales of $\sim100$ TeV can be probed.  It will be interesting to see if other possibilities for direct detection of light DM can be found.


\begin{theacknowledgments}
First, I would like to thank the SUSY '09 organizers for arranging such a pleasant and rewarding conference.  I would also like to gratefully thank A. Soni, my collaborator on the work on which this talk was based.  Additionally, I would like to thank M. Wise for his calculation of the decay width of $f\rightarrow e^+e^-\nu$ and for many helpful comments and suggestions, and also H. Davoudiasl, S. Dawson, S. Gopalakrishna, W. Marciano, C. Sturm, and M. Ramsey-Musolf for numerous helpful discussions.  This work is supported under US DOE contract No. DE-AC02-98CH10886.
\end{theacknowledgments}

\end{document}